\documentclass[submission,copyright,creativecommons]{eptcs}
\providecommand{\event}{GRAPHITE 2014} 

\newcommand{\algsize}{footnotesize}

\usepackage[ruled]{algorithm2e}
\usepackage{xspace}
\usepackage[usenames,dvipsnames]{color}
\usepackage{amsmath,amsthm}
\newtheorem{theorem}{Theorem}
\newtheorem{lemma}{Lemma}
\newtheorem{remark}{Remark}

\newcommand{\MDR}{\textsc{MinDR}}
\newcommand{\EMDR}{\textsc{ExtMinDR}}
\newcommand{\MIDR}{\textsc{MinDIR}}
\newcommand{\MCR}{\textsc{MaxCRS}}

\newcommand{\mytitle}{Entity-Linking via Graph-Distance Minimization}

\newcommand{\dfs}{\mathrm{dfs}}
\newcommand{\visited}{\mathrm{visited}}
\newcommand{\usefulEdgeFound}{\mathrm{usefulEdgeFound}}
\newcommand{\usefulEdge}{\mathrm{usefulEdge}}
\newcommand{\furthestAncestor}{\mathrm{furthestAncestor}}

\title{\mytitle}
\author{
Roi Blanco
\institute{Yahoo!~Research\\ Barcelona, Spain}
\email{roi@yahoo-inc.com}
\and
Paolo Boldi
\institute{Dipartimento di informatica\\ Universit\`a degli Studi di Milano}
\email{paolo.boldi@unimi.it}
\and
Andrea Marino\thanks{The second and third authors were supported by the EU-FET grant NADINE (GA 288956).}
\institute{Dipartimento di informatica\\ Universit\`a degli Studi di Milano}
\email{marino@di.unimi.it}
}
\def\titlerunning{\mytitle}
\def\authorrunning{R.~Blanco, P.~Boldi, A.~Marino}

\begin{document}

\maketitle

\begin{abstract}
Entity-linking is a natural-language--processing task that consists in 
identifying the entities mentioned in a piece of text, linking each to an appropriate
item in some knowledge base; when the knowledge base is Wikipedia, the problem comes
to be known as \emph{wikification} (in this case, items are wikipedia articles).
One instance of entity-linking can be formalized as an optimization
problem on the underlying concept graph, where the quantity to be optimized is the
average distance between chosen items.
Inspired by this application, we define a new graph problem which is a natural variant of 
the Maximum Capacity Representative Set. We prove that our problem is NP-hard for general graphs;
nonetheless, under some restrictive assumptions, it turns out to be solvable in linear time.
For the general case, we propose two heuristics: one tries to enforce the above assumptions
and another one is based on the notion of hitting distance; we show experimentally how these
approaches perform with respect to some baselines on a
real-world dataset.
\end{abstract}

\section{Introduction}

Wikipedia~\footnote{\url{http://en.wikipedia.org/}} is a free, collaborative, hypertextual
encyclopedia that aims at collecting articles on different (virtually, all)
branches of knowledge. The usage of wikipedia for automatically tagging
documents is a well-known methodology, that includes in particular a task called
\emph{wikification}~\cite{Mihalcea:2007}. Wikification is a special instance of
\emph{entity-linking}: a textual document is
given and within the document various fragments are identified (either manually
or automatically) as being \emph{(named) entities} (e.g., names of people,
brands, places\dots); the purpose of entity-linking is assigning a specific
reference (a wikipedia article, in the case of wikification) as a tag to each entity in 
the document.

Entity-linking happens typically in two stages: in a first phase, every entity is 
assigned to a set of items, e.g., wikipedia 
articles (the \emph{candidate nodes} for that entity); then a second phase consists
in selecting a single node for each entity, from within the set of candidates.
The latter task, called \emph{candidate selection}, is the topic on which this paper
focuses.

\medskip
To provide a concrete example, suppose that the target document contains 
the entity ``jaguar''
and the entity ``jungle''. Entity ``jaguar'' is assigned to a set of candidates
that contains (among others) both the wikipedia article about the feline living in America
and the one about the Jaguar car producer. On the other hand, ``jungle'' is
assigned to the article about tropical forests and to the one about the
electronic music genre. Actually, there are more than 30 candidates for
``jaguar'', and more about 20 for ``jungle''. 

In this paper, we study an instance of the candidate selection problem in which
the selection takes place based on some cost function that depends on the
average distance between the selected candidates, where the distance is measured
on the wikipedia graph\footnote{The undirected graph whose vertices are the
wikipedia articles and whose edges represent hyperlinks between them.}: the
rationale should be clear enough---concepts appearing in the same text are
related, and so we should choose, among the possible candidates for each entity,
those that are more closely related to one another. 

Getting back to the example above, there is an edge connecting ``jaguar'' the feline with 
``jungle'' the tropical forest, whereas the distance between, say, the feline and the music genre is
much larger.

\medskip
The approach we assume here highlights the \emph{collective} nature of the entity-linking
problem, as mentioned already in~\cite{HSZCELWTGM}: accuracy of the selection can be improved
by a global (rather than local) optimization of the choices. As~\cite{HSZCELWTGM} observes,
however, trying to optimize all-pair compatibility is a computationally difficult problem. 

In this paper, we prove that the problem itself, even in the simple instance we take into
consideration, is NP-hard; however, it becomes efficiently solvable under some special assumptions.
We prove that, although these assumptions fail to hold in real-world scenarios, we can still
provide heuristics to solve real instances. 

We test our proposals on a real-world dataset showing that
one of our heuristics is very effective, actually more effective than other
methods previously proposed in the literature, and more than a simple greedy
approach using the same cost function adopted here. 

\section{Related Work}

Named-entity linking (NEL)- also referred to as \emph{named entity
disambiguation} grounds mentions of entities in text (\emph{surface forms}) into
some knowledge base (e.g. Wikipedia, Freebase).
Early approaches to NEL~\cite{Mihalcea:2007} make use of measures derived from
the frequency of the keywords to be linked in the text and in different
Wikipedia pages. These include \emph{tf-idf}, $\chi^2$ and \emph{keyphraseness},
which stands for a measure of how much a certain word is used in Wikipedia links
in relation to its frequency in general text.
Cucerzan~\cite{Cucerzan07large-scalenamed} employed the context in which words
appears and Wikipedia page categories in order to create a richer representation
of the input text and candidate entities. These approaches were extended by
Milne and Witten~\cite{Milne:2008} who combined commonness (i.e., prior
probability) of an entity with its relatedness to the surrounding context using
machine learning. Further, Bunescu~\cite{bunescu08} employed a
\emph{disambiguation} kernel which uses the hierarchy of classes in Wikipedia
along with its word contents to derive a finer-grained similarity measure
between the candidate text and its context with the potential named entities to
link to.
In this paper we will make use of Kulkarni et al.'s
dataset~\cite{Kulkarni:2009}. They propose a general collective disambiguation
approach, under the premise that coherent documents refer to entities from one
or a few related topics. They introduce formulations that account for the
trade-off between local spot-to-entity compatibility and measures of global
coherence between entities.
More recently, Han et al.~\cite{HSZCELWTGM} propose a graph-based representation
which exploits the global interdependence of different linking decisions. The
algorithm infers jointly the disambiguated named mentions by exploiting the
graph.

It is worth to remark that NEL is a task somehow similar to Word Sense
Disambiguation (determining the right sense of a word given its context) in
which the role of the knowledge base is played by
Wordnet~\cite{Fellbaum-WordNet-1998}. WSD is a problem that has been extensively
studied and its explicitly connection with NEL was made by Hachey et
al~\cite{Hachey:2011}.
WSD has been an area of intense research in the past, so we will review here the
approaches that are directly relevant to our work.
Graph-based approaches to word sense disambiguation are pervasive and yield
state of the art performance~\cite{wsd}; however, its use for NEL has been
restricted to ranking candidate named entities with different flavors of
centrality measures, such as in-degree or PageRank~\cite{Hachey:2011}.

Mihalcea~\cite{Mihalcea:2005} introduced an unsupervised method for disambiguating the senses of words using random walks on graphs that encode the dependencies between word senses. 

Navigli and Lapata~\cite{Navigli:2005, Navigli:2007,Navigli:2010} present subsequent approaches to WSD using graph connectivity metrics, in which nodes are ranked with respect to their local \emph{importance}, which is regarded using centrality measures like in-degree, centrality, PageRank or HITS, among others.

Importantly, even if the experimental section of this paper deals with a NEL dataset exclusively, the theoretical findings could be equally applied to WSD-style problems.
Our \emph{greedy} algorithm is an adaptation of Navigli and Velardi's Structural Semantic Interconnections algorithms for WSD~\cite{Navigli:2005,Cuadros:2008}. The original algorithm receives an ordered list of words to disambiguate. The procedure first selects the \emph{unambiguous} words from the set (the ones with only one synset), and then for every ambiguous word, it iteratively selects the sense that is \emph{closer} to the sense of disambiguated words, and adds the word to the unambiguous set. This works in the case that a sufficiently connected amount of words is unambiguous; this is not the case in NEL and in our experimental set-up, where there could potentially exists hundreds of candidates for a particular piece of text. 

\section{Problem statement and NP-completeness}

In this section we will introduce the general formal definition of the problem, in the formulation
we decided to take into consideration.
We will make use of the classical graph notation: in particular, given an undirected graph
$G=(V,E)$, we will denote with $G[W]$ the graph induced by the vertices in $W$, and with
$d(u,v)$ the distance between the nodes $u$ and $v$, that is, the number of
edges in the shortest path from $u$ to $v$ (or the sum of the weights of the
lightest path, if $G$ is weighted). 

If $G$ is a graph and $e$ is an edge of $G$, $G-e$ is the graph obtained by removing $e$ from $G$; 
we say that $e$ is a \emph{bridge} if the number of connected components of $G-e$ is larger than that of
$G$. A connected bridgeless graph is called \emph{biconnected}; a maximal set of vertices of $G$ inducing
a biconnected subgraph is called a \emph{biconnected component} of $G$.

We call our main problem the \emph{Minimum Distance Representative}, in short
\MDR, and we define it as follows.
Given an undirected graph $G=(V,E)$ (possibly weighted) and $k$ subsets of its
set of vertices, $X_1,\ldots, X_k\subseteq V$, a feasible solution for 
\MDR\ is a sequence of vertices of $G$, $x_1,\ldots,x_k$, such that for any $i$, with $1\leq i \leq k$, $x_i\in X_i$ (i.e., the solution 
contains exactly one element from every set, possibly with repetitions).

Given the instance $G,
\{X_1,\ldots, X_k\}$, the measure (the \emph{distance cost}) of a solution $S$, $x_1,\ldots, x_k$, is 
$f(S)=\sum_{i=1}^k\sum_{j=1}^k d(x_i,x_j)$. The goal
is finding the solution of minimum distance cost, i.e., a feasible solution $S$ such that
$f(S)$ is minimum.

We call the restriction of this problem, in which the sets of vertices in input $\{X_1,\ldots , X_k\}$ are disjoint, 
\MIDR (Minimum Independent Distance Representative). In this case, for the sake of simplicity, we will refer to a solution as the multiset composed by its elements.\footnote{We shall make free use of multiset membership, intersection and union with their 
standard meaning: in particular, if $A$ and $B$ are multisets
with multiplicity function $a$ and $b$, respectively, the multiplicity functions of 
$A \cup B$ and $A \cap B$ are $x \mapsto \max(a(x),b(x))$ and $x \mapsto \min(a(x),b(x))$, respectively.}

\subsection{NP-completeness of \MDR}

The \MIDR\ problem seems to be similar and related to the so-called Maximum Capacity Representatives 
\cite{CrescenziK97}, in short \MCR. The Maximum Capacity Representatives problem is defined as follows: 
given some disjoint sets $X_1,\ldots , X_m$ and for any $i \neq j$, $x\in X_i$, and $y\in X_j$, a nonnegative 
capacity $c(x,y)$, a solution is a set $S=\{x_1,\ldots x_m\}$, such that, for any $i$, $x_i\in X_i$; such a 
solution is called \emph{system of representatives}. The measure of a solution is the capacity of the system of representatives, 
that is $\sum_{x\in S}\sum_{y\in S}c(x,y)$, and the \MCR\ problem aims at \emph{maximizing} it. 
The \MCR\ problem was introduced by \cite{Bellare93}, who showed that it is 
NP-complete and gave some non-approximability results. Successively, in \cite{Serna2005123}, tight inapproximability 
results for the problem were presented.

The \MIDR\ problem differs from \MCR\ just for in the sense that we are dealing
with distances instead of capacities, and therefore we
ask for a minimum instead of a maximum. Nonetheless the following Lemma, whose proof is given in Appendix \ref{app:one}, shows that also \MIDR\ problem is NP-complete.

\begin{lemma}
The \MIDR\ (hence, \MDR) problem is NP-complete.
\label{lem:np}
\end{lemma}

\section{The decomposable case}

In this section we study the \MDR\ problem under some restrictive
hypothesis and we will show that in this case a linear exact
algorithm exists. 

Even if it may seem that these hypothesis are too strong to make the algorithm
useful in practice, in the next section we will use our algorithm to design
an effective heuristic for the general problem. In particular, we assume that the
graph $G$ (possibly weighted) is such that:
\begin{itemize}
\item any set $X_i$ induces a connected subgraph on $G$, i.e.,  $G[X_i]$ is connected,
\item for any $i \neq j$, for any $x\in X_i$ and $y\in X_j$, $x$ and $y$ do not belong to the same biconnected component.
\end{itemize}
The problem, under these further restrictions, will be called \emph{decomposable \MDR}.
Note that the second condition implies that a decomposable \MDR\ is in fact an instance of \MIDR, because
it implies that no two sets can have nonempty intersection.

Let us consider an instance $(G,\{X_1,\ldots X_k\})$ of decomposable \MDR\
problem on a graph $G=(V,E)$. 

An edge $e=(x,y) \in E$ is called \emph{useful} if it is a bridge, $x$ and $y$ do not belong to the same set 
$X_i$, and there are at least two indices $i$ and $j$ such that $X_i$ and $X_j$ are in different components of 
$G-e$ (since $e$ is a bridge, the graph obtained removing the edge $e$ from $G$ is no more connected).

\subsection{Decomposing the problem}

The main trick that allows to obtain a linear-time solution for the decomposable case is that we can actually decompose
the problem (hence the name) through useful edges. First observe that, trivially:

\begin{remark}
\label{rem:bridge}
Let $e=(x,y)$ be a useful edge and let $Z_x$ and $Z_y$ be the two connected
components of $G-e$ containing $x$ and $y$, respectively. In $G$, all paths from
any $x' \in Z_x$ to any $y' \in Z_y$ must contain $e$.
\end{remark}

Moreover:

\begin{remark}
\label{rem:dontsplit}
Let $e=(x,y)$ be a useful edge. There cannot be an index $i$ such that $X_i$ has a nonempty intersection
with both components of $G-e$.
\end{remark}

In fact, assume by contradiction that one such $X_i$ exists, and let $u, w \in X_i$ be two vertices living in
the two different components of $G-e$: since $G[X_i]$ is connected, there must be a path connecting $u$ and $w$ and
made only of elements of $X_i$; because of Remark~\ref{rem:bridge}, this path passes through $e$, but this would
imply that $x,y \in X_i$, in contrast with the definition of useful edge.

\medskip
Armed with the previous observations, we can give the following further definitions.
Let $Y_x$ (respectively, $Y_y$) be the set of sets $X_i$ such that
$X_i \subseteq Z_x$ (respectively, $X_i \subseteq Z_y$); we denote the sets of
nodes in $Y_x$ and $Y_y$ by $V(Y_x) \subseteq Z_x$ and $V(Y_y)
\subseteq Z_y$, respectively.

By virtue of Remark~\ref{rem:bridge}, all the paths in $G$ from any $x' \in V(Y_x)$ to any $y' \in V(Y_y)$ pass through $e$.
This implies also that there is no simple cycle in the graph including both $x' \in V(Y_x)$ and $y' \in V(Y_y)$.

Given a solution $S$ for \MIDR$(G,\{X_1,\ldots, X_k\})$, and a useful edge $(x,y)$, we have:
\begin{eqnarray*}
\sum_{x_i, x_j\in S} d(x_i, x_j) 	& =	& \sum_{x_i,x_j\in S\cap V(Y_x)} d(x_i,x_j) +\sum_{x_i,x_j\in S\cap V(Y_y)} d(x_i,x_j) + \\
													& 		&2\sum_{x_i\in S\cap V(Y_x) , x_j\in S\cap V(Y_y)} \left(d(x_i, x) + d(x,y) + d(y, x_j) \right).
\end{eqnarray*}
Indeed all the shortest paths from any $x_i\in S\cap V(Y_x)$ to any $x_j\in S\cap V(Y_y)$ pass through the useful edge $(x,y)$ by Remark \ref{rem:bridge}. 
Moreover, since the sets $X_1,\ldots, X_k$ are disjoint, we have that $|S\cap
V(Y_x)|=|Y_x|$ and $|S\cap V(Y_y)|=|Y_y|$, that is, a solution has exactly
one element for each set in $Y_x$ (respectively, $Y_y$). Hence we can rewrite the
last summand of the above equation as follows:
\begin{eqnarray*}
\sum_{x_i\in S\cap V(Y_x) , x_j\in S\cap V(Y_y)} \left( d(x_i, x) + d(y, x_j) +d(x,y)\right)  	& = 	
&         |Y_y| \cdot \sum_{x_i\in S\cap V(Y_x)} d(x_i,x) + \\
& 		& |Y_x| \cdot \sum_{x_j\in S\cap V(Y_y)} d(y, x_j) +\\
& 		& |Y_x| \cdot |Y_y|\cdot  d(x,y). 
\end{eqnarray*}

By combining the two equations, we can conclude that finding a solution for
\MIDR$(G,\{X_1,\ldots, X_k\})$ can be decomposed into the
following two subproblems:
\begin{enumerate}
\item finding $S_x$ minimizing $\sum_{x_i,x_j\in S\cap V(Y_x)}
d(x_i,x_j)+2\sum_{x_i\in S\cap V(Y_x)} |Y_y|d(x_i,x)$ in the instance
$(G[Z_x],Y_x)$;
\item finding $S_y$ minimizing $\sum_{x_i,x_j\in S\cap V(Y_y)}
d(x_i,x_j)+2\sum_{x_j\in S\cap V(Y_y)} |Y_x|d(y,x_j)$ in the instance
$(G[Z_y],Y_y)$.
\end{enumerate}

Note that both instances are smaller than the original one because of the
definition of a useful edge. The idea of our algorithm generalizes this
principle; note that the new objective function we must take into consideration
is slightly more complex than the original one: in fact, besides the usual
all-pair--distance cost there is a further summand that is a weighted sum of
distances from some fixed nodes (such as $x$ for the instance $G[Z_x],Y_x$ and
$y$ for the instance $G[Z_y],Y_y$).

We hence define an extension of the \MDR\ problem, that we call \EMDR\ (for
\emph{Extended Minimum Distance Representatives}).
In this problem, we are given:
\begin{itemize}
  \item an undirected graph $G=(V,E)$ (possibly weighted)
  \item $k$ subsets of its set of vertices, $X_1,\ldots, X_k\subseteq V$
  \item a multiset $B$ of vertices, each $x \in B$ endowed with a weight $b(x)$.
\end{itemize}
A feasible solution for the \EMDR\ is a multiset $S=\{x_1,\ldots,x_k\}$ of vertices
of $G$, such that for any $i$, with $1\leq i \leq k$, $S\cap X_i\neq \emptyset$ (i.e., the set contains at least
one element from every set). Its cost is 
\[
	f(S)=\sum_{i=1}^h\sum_{j=1}^k d(x_i,x_j)+\sum_{i=1}^k\sum_{z \in B}
	b(z)d(x_i,z).
\]
The goal is finding the solution of minimum cost, i.e., a feasible solution $S$
such that $f(S)$ is minimum. The original version of the problem is obtained by letting $B=\emptyset$.

\medskip
We are now ready to formalize our decomposition through the following Theorem, whose proof is given in Appendix \ref{app:two}.
\begin{theorem}
\label{thm:decomposition}
Let us be given a decomposable \EMDR\ instance $(G,\{X_1,\dots,X_k\},B,b)$
and a useful edge $e=(t_0,t_1)$. 
For every $s \in \{0,1\}$, let $Z_s$ be the connected component of $G-e$ containing $t_s$, $Y_s$ be the set
of sets $X_i$ such that $X_i \subseteq Z_s$ and $V(Y_s)$ be the union of those
$X_i$'s. Let also $B_s$ be the intersection of $B$ with $Z_s$. 
Define a new instance $I_s=(T[Z_s],\{X_i, i \in Y_s\},B_s\cup\{t_s\},b_s)$ where
\[
	b_s(t_s)=2|Y_{1-s}|+\sum_{z \in B_{1-s}}b(z)\textrm{\ and \ }
	b_s(z)=b(z),\textrm{ for any $z\in B$.}
\]
Then the cost $f(S)$ of an optimal solution $S$ of the original problem is equal
to 
$$f(S_0)+f(S_1)+2|Y_0||Y_1|d(t_0,t_1) + \sum_{s\in\{0,1\}}\left(|S\cap V(Y_s)|\cdot \sum_{z\in B\cap Z_{1-s}}b(z)d(t_s,z)\right)$$ 
where $S_s$ is an optimal solution for the instance $I_s$.
\end{theorem}

For completeness, we need to consider the base case of
an instance with just one set $G, \{X_1\}, B, b$: the
solution in this case is just one node $x \in X_1$ and the objective function to be minimized is
simply $\sum_{z\in B}d(x,z)b(z)$. The optimal solution can be found by
performing a BFS from every $z_j\in B$ (in increasing order of $j$), maintaining for each node $y\in X_1$,
$g(y)=\sum_{z_t\in B, t<j}d(x,z_t)b(z_t)$, and picking the node having maximum
final $g(y)$. This process takes $O(|B|\cdot |E(G[X_1])|)$.
It is worth observing that in our case the size of the multiset $B$ is always bounded by $k$. 
Moreover since $\sum_{i=1}^k |E(G[X_i])| \leq |E(G)|=m$, the overall complexity for all these base cases is 
bounded by $O(k\cdot m)$.

\subsection{Finding useful edges}

For every instance with more than one set, given an useful edge $e$ the
creation of the subproblems as described above is linear, so we are left with the issue of finding useful edges.
This task can be seen as a variant of the standard depth-first search of bridges, as
shown in Algorithm~\ref{alg:useful} and~\ref{alg:useful-dfs}, in Appendix \ref{app:three}. 
Recall that bridges can be found by performing a standard DFS that numbers the nodes as they
are found (using the global counter $\visited$, and keeping the DFS numbers in the array $\dfs$);
every visit returns the index of the least ancestor reachable through a back edge while visiting
the DFS-subtree rooted at the node where the visit starts from.
Every time a DFS returns a value that is larger than the number of the node currently being
visited, we have found a bridge.

The variant consists in returning not just the index of the least ancestor reachable, but
also the set of indices $i$ that are found while visiting the subtree. If the set of indices
and its complement are both different from $\emptyset$ then the bridge is useful:
at this point, a ``rapid ascent'' is performed to get out of the recursive procedure.

\subsection{The final algorithm}

Combining the observations above, 
we can conclude that the overall complexity of the algorithm is $O(k\cdot m)$.
The algorithm is presented in Algorithm~\ref{alg:one}.

\begin{algorithm}[t]
\begin{\algsize}
\KwIn{A graph $G=(V,E)$, $X_1\ldots,X_k\subseteq V$, a weighted multiset $B$ of nodes in $V$, where  each element in $B$ has a weight $b$. 
$G[X_i]$ is connected for every $i$ and moreover for all $i\neq j$ and $x \in X_i$, $y \in X_j$, the two vertices $x$ and $y$ do not
belong to the same biconnected component of $G$.}
\KwOut{A solution $S=\{x_1, \ldots, x_k\}$ such that for any $i$, with $1\leq i \leq k$, $x_i \in X_i$, minimizing 
$\sum_{i=1}^h\sum_{j=1}^k d(x_i,x_j)+\sum_{i=1}^k\sum_{z \in B}
	b(z)d(x_i,z)$}
Find a useful edge $e=(x,y)$, if it exists, using Algorithm~\ref{alg:useful}\\
\eIf{the useful edge does not exist}{
	\If{$k\neq 1$}{
		Fail! 
	}
	Output the element $x_1 \in X_1$ minimizing $\sum_{z \in B}b(z)d(x_1,z)$\\ 
}{
	Let $Z_x$ (respectively $Z_y$) be the connected component of $T-e$ containing
	$x$ (respectively $y$) . \\
	Let $Y_x$ (respectively $Y_y$) be the indices $i$ such that $X_i\subseteq Y_x$
	($X_i\subseteq Y_y$, respectively)\\
	$B'\gets B\cup \{x\}$ (multiset union) with $b(x)=2|Y_y|+\sum_{z\in B\cap Z_y}b(z)$\\
	$B'\gets B'\cap Z_x$ (multiset intersection)\\
	$S'\gets \mathrm{\textsc{DecomposableMinDR}}(T[Z_x],Y_x,B')$\\
	$B''\gets B\cup \{x\}$ (multiset union) with $b(y)=2|Y_x|+\sum_{z\in B\cap Z_x}b(z)$\\
	$B''\gets B''\cap Z_y$ (multiset intersection)\\
	$S''\gets \mathrm{\textsc{DecomposableMinDR}}(T[Z_y],Y_y,B'')$\\
	\Return{$S'\cup S''$} 
}
\end{\algsize}
\caption{\textsc{DecomposableMinDR}}
\label{alg:one}
\end{algorithm}

\section{The general case}
\label{sec:heuristics}

As we observed at the beginning, the \MDR\ problem is NP-complete in general, although the decomposable
version turns out to be linear. We want to discuss how we can deal with a general instance of the problem.
To start with, let us consider a general connected 
\MDR\ instance, that is:
\begin{itemize}
  \item a connected undirected (possibly weighted) graph $G=(V,E)$,
  \item $k$ subsets of its set of vertices, $X_1,\ldots, X_k\subseteq V$,
\end{itemize}
with the additional assumption that $G[X_i]$ is connected for every $i$.
Recall that a feasible solution is a sequence $S$ of vertices of $G$, $ x_1,\ldots,x_k$, such that for any $i$, with 
$1\leq i \leq k$, we have $x_i\in X_i$;
its (distance) cost is $f(S)=\sum_{i=1}^k\sum_{j=1}^k d(x_i,x_j)$.

We shall discuss two heuristics to approach this problem: the first is related to Algorithm~\ref{alg:one} in that it tries
to modify the problem to make it into a decomposable one, whereas the second is based on the notion of hitting
distance. 

Before describing the two heuristics, let us briefly explain the rationale behind the additional assumption (i.e., that every $G[X_i]$ be connected).
In our main application (entity-linking) the structure of the graph within each $X_i$ is not very important, and can actually be misleading:
a very central node in a large candidate set may seem very promising (and may actually minimize the distance to the other sets) but can be
blatantly wrong. It is pretty much like the distinction between nepotistic and non-nepotistic links in PageRank computation: the links
\emph{within} each host are not very useful in determining the importance of a page---on the contrary, they may be confusing, and are thus
often disregarded.

Based on this observation, we can (and probably want to) modify the structure of the graph within each set $X_i$ to avoid this kind
of trap. This is done by preserving the \emph{external} links (those that connect vertices of $X_i$ to the outside), but
at the same time adding or deleting edges within each $X_i$ in a suitable way. In our experiments, we considered 
two possible approaches:
\begin{itemize}
  \item one consists in making $G[X_i]$ \emph{maximally connected}, i.e., transforming it into a clique;
  \item the opposite approach makes $G[X_i]$ \emph{minimally connected} by adding the minimum number of edges needed to that purpose;
  this can be done by computing the connected components of $G[X_i]$ and then adding enough edges to join them in a single connected component.
\end{itemize} 
Both approaches guarantee that $G[X_i]$ is connected, so that the two heuristics described below can be applied.

\subsection{The spanning-tree heuristic}
The first heuristic aims at modifying the graph $G$ in such a way that the resulting instance becomes decomposable.
For the moment, let us assume that the sets $X_i$ are pairwise disjoint.
To guarantee that the problem be decomposable, we proceed as follows. Define an equivalence relation $\sim$
on $V$ by letting $x \sim y$ whenever $x$ and $y$ belong to the same $X_i$.\footnote{Note that, since the sets $X_i$ are pairwise disjoint, $\sim$ is transitive.}
The quotient graph $G/\sim=(V/\sim,E/\sim)$ has vertices $V/\sim$ and an edge between $[x]$ and $[y]$ whenever there
is some edge $(x',y') \in E$ with $x' \sim x$ and $y' \sim y$ (here, and in the following, $[x]$ denotes the $\sim$-equivalence class including $x$).
Thus, there is a surjective (but not injective) map $\iota: E \to E/\sim$.

Since $G$ is connected, so is $G/\sim$, and we perform a breadth-first traversal of $G$ building a spanning tree $T$.
Every tree edge is an edge of $G/\sim$, so its pre-image with respect to $\iota$ is a nonempty
set of edges in $G$. Let us arbitrarily choose one edge of $G$ from $\iota^{-1}(t)$ for every tree edge $t$,
and let $T'$ be the resulting set of edges of $G$.
 
Define the new graph $G'=(V,E')$ where $E'=T' \cup \bigcup_{i=1}^k E(G[X_i])$: this graph cointains all
the edges within each set $X_i$, plus the set $T'$ of external edges. 

It is easy to see that $G'[X_i]$ is connected (it is in fact equal to $G[X_i]$), and moreover all the elements of
$T'$ are bridges dividing all the $X_i$'s in distinct biconnected components.
In other words, we have turned the instance into a \emph{decomposable} one, where Algorithm~\ref{alg:one} can be run. 

\paragraph{The non-disjoint case}
If the sets $X_i$ are not pairwise disjoint, we can proceed as follows.
Let us define maximal mutually disjoint sets of indices $I_1, \dots, I_h \subseteq\{1, \dots, k\}$
such that for all $t\neq s$, $\cup_{i \in I_t} X_i \cap \cup_{i \in I_s} X_i = \emptyset$.

Now, take the new problem instance with the same graph and sets $Y_1,\dots,Y_h$ where
$Y_t=\cup_{i \in I_t} X_i$: this instance is disjoint, so the previous construction applies.
The only difference is that, at the very last step of Algorithm~\ref{alg:one}, when we are left
with a graph and a \emph{single} $Y_t$, we will not select a single $y \in Y_t$ optimizing the
cost function
\[
	\sum_{z \in B}b(z)d(y,z).
\]
Rather, we will choose one element $x_i$ for every $i \in I_t$ optimizing
\[
	\sum_{i \in I_t}\sum_{z \in B}b(z)d(x_i,z).
\]

\paragraph{Discussion}
Both steps presented above introduce some level of imprecision, that make the algorithm only
a heuristic in the general case. The first approximation is due to the fact that building a tree
on $G$ will produce distances (between vertices living in different $X_i$) much larger than they are in $G$;
the second approximation is that when we have non-disjoint sets, we only optimize with respect to 
bridges, disregarding the sum of distances of the nodes of different sets. Actually, we should
optimize 
\[
	\sum_{i \in I_t}\sum{j \in I_t} d(x_i,x_j)+\sum_{i \in I_t}\sum_{z \in B}b(z)d(x_i,z).
\]
but this would make the final optimization step NP-complete. 

\subsection{The hitting-distance heuristic}
The second heuristic we propose is based on the notion of \emph{hitting distance}: given a vertex $x$ and a set of vertices
$Y$, define the hitting distance of $x$ to $Y$ as $d(x,Y)=\min_{y \in Y} d(x,y)$. The hitting distance can be easily found 
by a breadth-first traversal starting at $x$ and stopping as soon as an element of $Y$ is hit. 
Given a general connected instance of \MDR, as described above, we can consider, for every $i$ and every $x \in X_i$, the 
average hitting distance of $x$ to the other sets:
\[
	\frac{\sum_{j=1}^k d(x,X_j)}{k}.
\] 
The element $x_i^* \in X_i$ minimizing the average hitting distance (or any such an element, if there are many) is the candidate
chosen for the set $X_i$ in that solution. 

The main problem with this heuristic is related to its locality (optimization is performed separately for each $X_i$); moreover
the worst-case complexity is $O(m \sum_i |X_i|)$, that reduces to $O(k \cdot m)$ only under the restriction that the 
sets $X_i$ have $O(1)$ size.

\section{Experiments}

All our experiments were performed on a snapshot of the English portion of
Wikipedia as of late February 2013; the graph (represented in the BVGraph format~\cite{BoVWFI}) was symmetrized and only the largest component was kept. 
The undirected graph has 3\,685\,351 vertices ($87.2\%$ of the vertices of
the original graph) and 36\,066\,162 edges ($99.9\%$ of the edges of the
original graph).
Such a graph will be called the ``Wikipedia graph'' and referred to as $G$ throughout this experimental section.
 
Our experiments use actual real-world entity-linking problems for which we have a
human judgment, and tries the two heuristics proposed in
Section~\ref{sec:heuristics}, as well as a greedy baseline and other heuristics.

The greedy baseline works as follows: it first chooses an index $i$ at random, and draws an element $x_i \in X_i$ also at random.
Then, it selects a vertex of $x_{i+1} \in X_{i+1},x_{i+2} \in X_{i+2},\dots,x_k \in X_k, x_1 \in X_1, \dots, x_{i-1} \in X_{i-1}$ 
(in this order) minimizing each time the sum of the distances
to the previously selected vertices; the greedy algorithm continues doing the same also for $x_i \in X_i$ to
get rid of the only element (the first one) that was selected completely at
random. Moreover we have considered also two other heuristics, that have been
observed to be effective in practice~\cite{Hachey:2011}: these are
\emph{degree} and \emph{PageRank based}. They respectively select the highest
degree and the highest PageRank vertex for each set. 

The real-world entity-linking dataset has been taken from~\cite{Kulkarni:2009}
which contains a larger number of human-labelled annotations. For retrieving the
candidates, we created an index over all Wikipedia pages with different fields
(title, body, anchor text) and used a variant of BM25F~\cite{Blanco:2012} for
ranking, returning the top 100 scoring candidate entities. Since the
candidate selection method was the same for every graph-based method employed,
there should be no bias in the experimental outcomes.


The problem instances contained in the dataset have $11.73$ entities on average
(with a maximum of $53$), and the average number of candidates per entity is
$95.90$ (with a maximum of $200$).
Each of the 100 problem instances in the NEL dataset is annotated, and in
particular, for every $i$ there is a subset $X_i^* \subseteq X_i$ of \emph{fair} vertices (that is, vertices that are good candidates for that set): typically $|X_i^*|=1$.
Note that, for every instance in the NEL dataset, we deleted the sets $X_i$ such that $X_i^*$ were not included in the largest connected
component of the Wikipedia graph. The number of sets $X_i$ deleted was at maximum 2 (for two instances). We have not considered instances in which, after these modifications, 
we have just one set $X_i$: this situation happened in 5 cases. So the problem
set on which we actually ran our algorithm contains 95 instances.


For every instance, we considered the maximal and
minimal connection~\footnote{To obtain the minimal connection of each $G[X_i]$, we chose to connect the vertex of maximum degree of its largest component with an (arbitrary) vertex of each of its remaining components.} 
approach, and
then ran both heuristics described in Section~\ref{sec:heuristics}, comparing them with the greedy baseline, and also with the degree and PageRank heuristics.

For any instance, when comparing the distance cost $f$ of the solutions $S_j$
returned by some algorithm $A_j$, we have computed the \emph{distance-cost
ratio} of each algorithm $A_j$, defined as
$$\frac{f(S_j)}{\min_{j} f(S_j) }\cdot 100.$$ Intuitively this corresponds to
the approximation ratio of each solution with respect to the best solution found
by all the considered algorithms: hence the best algorithm has minimum
distance-cost ratio and it equals 100.

Besides evaluating the distance cost of the solutions found by the various
heuristics, we can compute how many of the elements found are fair: we normalize
this quantity by $k$, so that $1.0$ means that all the $k$ candidates selected
are fair. We call such a quantity the \emph{value} of a solution.

\begin{table}[t]
\begin{small}
\centering
\begin{tabular}{|l||c|c||c|c|}
\hline
& \multicolumn{2}{c||}{\textsc{Distance-cost ratio}} &
\multicolumn{2}{c|}{\textsc{Value}}\\
\cline{2-5}
& \multicolumn{1}{c|}{\textsc{Maximal}} & \multicolumn{1}{c||}{\textsc{Minimal}}& \multicolumn{1}{c|}{\textsc{Maximal}} & \multicolumn{1}{c|}{\textsc{Minimal}}\\
& \multicolumn{1}{c|}{\textsc{Connection}} & \multicolumn{1}{c||}{\textsc{Connection}}& \multicolumn{1}{c|}{\textsc{Connection}} & \multicolumn{1}{c|}{\textsc{Connection}}\\
\cline{2-5}
& Average & Average & Average & Average\\
\textsc{Heuristic} & ($\pm$ Std Error) &($\pm$ Std Error) & ($\pm$ Std Error) &($\pm$ Std Error)\\
\hline
Spanning-tree & 122.747($\pm$2.812) & 130.998 ($\pm$2.917)& 0.369 ($\pm$0.023) &
0.360 ($\pm$0.023)\\
Hitting-distance & 103.945 ($\pm$1.320) & 105.797 
($\pm$2.322) & {\bf 0.454 ($\pm$0.027)} & {\bf 0.459 ($\pm$0.027)}\\
\hline
Greedy & {\bf 101.969 ($\pm$0.429)} & {\bf 102.785 ($\pm$ 0.426)} & 0.428
($\pm$0.025) & 0.426 ($\pm$0.026)\\
Degree based &  114.182 ($\pm$2.386) & 113.285 ($\pm$2.305) & 0.411
($\pm$0.024) & 0.394 ($\pm$0.023)\\
PageRank based & 114.894 ($\pm$2.452) & 112.392 ($\pm$2.266) & 0.407 ($\pm$0.025) & 0.398 ($\pm$0.023)\\
\hline
\hline
\textsc{Ground truth} & 115.117 ($\pm$1.782) & 119.243 ($\pm$1.873)\\
\cline{1-3}
\end{tabular}
\caption{Distance-cost ratio and value.}
\label{tab:ratio}
\end{small}
\end{table}

In the last two columns of Table \ref{tab:ratio} we report, for each heuristic,
the average value (across all the instances)
along with the standard error. For both the connection
approaches, we have that the hitting-distance heuristic outperforms all the
other heuristics, and it selects more than 45\% of fair candidates. The
variability of the results seems not to differ too much for all the methods. The
second best heuristic is the greedy baseline, that selects almost 42.8\% and
42.6\% fair candidates respectively in a maximal and minimal connected scenario.

It is worth observing that the greedy approach comes second (as far as the
value is concerned), and outperforms the baseline techniques (degree and
PageRank). The spanning tree heuristic, instead, perform worse than any other
method.

The latter outcome is easily explained by the fact that
it transforms completely the topology of the graph in order to make the
instance decomposable, and the distances between vertices are mostly scrambled.
This interpretation of the bad result obtained can also be seen looking at the 
distance cost (central columns of Table~\ref{tab:ratio}): the spanning-tree
heuristic is the one that is less respectful of distances, selecting candidates
that are far apart from one another.

In the central columns of Table \ref{tab:ratio}, we report also the
distance-cost ratio for all the other heuristics. For both the maximal and the
minimal connection approaches, the greedy baseline seems to obtain more often a
minimum distance cost solution. The second best option is the hitting distance
heuristic, while the other methods seems to be more far away from an optimal
result.

In the last row of Table \ref{tab:ratio}, we
report the distance-cost ratio for the ground-truth solution given by the fair
candidates.
It seems that for any instance, the ground truth has distance cost averagely
15\%-20\% higher than the best solution we achieve by using the heuristics. 
This observation suggests that probably our objective function (that simply
aims at minimizing the graph distances) is too simplistic: the distance cost is
an important factor to be taken into account but certainly not the unique one. 

It is interesting to remark, though, that
the average Jaccard coefficient between the solution found by the
degree based and the hitting-distance heuristic is 0.3 (for both
maximal and minimal connection approaches): this fact means that the degree and
distance can be probably used as complementary features that hint at different 
good candidates, although we currently do not know how to combine these
pieces of information.

Finally, we remark that we also tried to apply the degree and PageRank based
heuristics by using the same problem set but \emph{in the original directed
graph}; in this case, we did not enforce any connectivity of the subgraphs
$G[X_i]$:
the resulting average values ($\pm$ standard error) are respectively $0.327$
($\pm 0.020$) and $0.336$ ($\pm 0.022$), and they are both worse than the values
achieved by degree and PageRank heuristics in Table \ref{tab:ratio}. This fact
suggests that our experimental approach (of considering the undirected version
and of enforcing some connectivity on the subgraphs) not only guarantees the
applicability of our heuristics in a more suitable scenario, but also improves
the effectiveness of the other existing techniques.

\section{Conclusions and future work} 
 
Inspired by the entity-linking task in NLP, we defined and studied a new graph
problem related to Maximum Capacity Representative Set and we proved that
this problem is NP-hard in general (although it remains an open problem to
determine its exact approximability).
Morevoer, we showed that the problem can be solved efficiently in
some special case, and that we can anyway provide reasonable heuristics for the
general scenario. We tested our proposals on a real-world dataset showing that
one of our heuristics is very effective, actually more effective than other
methods previously proposed in the literature, and more than a simple greedy
approach using the same cost function adopted here.

The other heuristic proposed in this paper seem to work poorly (albeit it
reduces to a case where we know how to produce the optimal solution), but we
believe that this is just because of the very rough preprocessing phase it
adopts; we plan to devise a more refined way to induce the conditions needed for
Algorithm~\ref{alg:one} to work, without having to resort to the usage of a
spanning tree---the latter scrambles the distances too much, resulting in a bad
selection of candidates.

Finally, we observed that a distance-based approach is complementary to other
methods (e.g., the local techniques based solely on the vertex degree), hinting
at the possibility of obtaining a new, better cost function that exploits
both features at the same time.

\bibliographystyle{eptcs}
\bibliography{main}

\newpage
\appendix

\section{Proof of Lemma \ref{lem:np}}
\label{app:one}
\begin{proof}
We reduce \MCR\ to \MIDR. Given an instance of \MCR, $\{X_1,\ldots X_k\}$ and
for any $i \neq j$, $x\in X_i$, and $y\in X_j$, a nonnegative capacity $c(x,y)$,
we construct the instance of \MIDR\ $G,\{X_1,\ldots, X_k\}$;
the vertices of $G$ are $X_1\cup \ldots \cup X_k$, and for any pair $x\in
X_i$, $y\in X_j$, with $i\neq j$, we add a weighted edge between $x$ and $y$,
i.e., for each pair for which \MCR\ defines a capacity we create a corresponding
edge in $G$. In particular the weight of the edge between $x$ and $y$ is set to
$\alpha - c(x,y)$, where $\alpha=2\max_{z\in X_i, t\in X_j, i\neq j} c(z,t)$.

Observe that for any pair of nodes $u\in X_i$, $v\in X_j$, with $i\neq j$,
$d(u,v)$ in $G$ is equal to the weight of $(u,v)$, i.e., it is not convenient to
pass through other nodes when going from $u$ to $v$: in fact, for any path $z_1,\ldots, z_p$ from $u$ to $v$
in $G$, with $p\geq 1$, we always have $\alpha - c(u,v)\leq \alpha-c(u,z_1) +
\ldots + \alpha-c(z_p,v)$, since $\alpha - c(u,v)\leq \alpha$ and the weight of
such a path is at least $\frac{p+1}{2}\alpha\geq \alpha$. Moreover, observe that
any optimal solution in $G$ has exactly one element for each set $X_i$: thus, we
have $k(k-1)$ pairs of elements $(x,y)$, whose distance is always given by the
weight of the single edge $(x,y)$, that is $\alpha-c(x,y)$.

Hence it is easy to see that \MCR\ admits a system of representatives whose
capacity is greater than $h$, if and only if \MIDR\ admits a solution $S$ such
that $f(S)$ is less than $k(k-1)\alpha - h$.

Since \MIDR\ is a restriction of \MDR\, we can conclude that also \MDR\ is NP-complete.
\end{proof}

\section{Proof of Theorem \ref{thm:decomposition}}
\label{app:two}
\begin{proof}
We can rewrite the objective function as follows.
\begin{eqnarray*}
\sum_{x_i, x_j\in S} d(x_i, x_j)+\sum_{x_i\in S}\sum_{z\in B}d(x_i,z)b(z)
&=&  2|Y_0| |Y_1|d(t_0,t_1) + \sum_{x_i,x_j\in S\cap V(Y_0)} d(x_i,x_j) + \sum_{x_i,x_j\in S\cap V(Y_1)} d(x_i,x_j) +\\
& & 2|Y_1|\sum_{x_i\in S\cap V(Y_0)}
d(x_i,t_0) + \sum_{x_i\in S\cap V(Y_0)}\sum_{z\in B}d(x_i,z)b(z)+\\
& & 2|Y_0|\sum_{x_j\in S\cap V(Y_1)}
d(t_1, x_j) + \sum_{x_i\in S\cap V(Y_1)}\sum_{z\in B}d(x_i,z)b(z).
\end{eqnarray*}
This is because if $z\in B\cap Z_1$, for any node $x_i\in S\cap V(Y_0)$, we have 
$d(x_i,z)=d(x_i,t_0)+d(t_0,z)$ (and analogously, if $z\in B\cap Z_0$, for any
node $x_i\in S\cap V(Y_1)$, we have $d(x_i,z)=d(x_i,t_1)+d(t_1,z)$). Hence:
\begin{eqnarray*}
\sum_{x_i\in S\cap V(Y_0)}\sum_{z\in B}d(x_i,z)b(z) &=& \sum_{x_i\in S\cap
V(Y_0)}\sum_{z\in B\cap Z_0}  d(x_i,z)b(z) + \sum_{x_i\in S\cap
V(Y_0)}\sum_{z\in B\cap Z_1} d(x_i,t_0)b(z)+d(t_0,z)b(z)
\end{eqnarray*}
and
\begin{eqnarray*}
\sum_{x_i\in S\cap V(Y_1)}\sum_{z\in B}d(x_i,z)b(z) &=& \sum_{x_i\in S\cap
V(Y_1)}\sum_{z\in B\cap Z_1}  d(x_i,z)b(z) + \sum_{x_i\in S\cap
V(Y_1)}\sum_{z\in B\cap Z_0} d(x_i,t_1)b(z)+d(t_1,z)b(z).
\end{eqnarray*}
Observe that $t_0$ or $t_1$ might already belong to $B$: this is why we assumed
that $B$ is a multiset. 

Then, we have that:


$$f(S_0)=\sum_{x_i,x_j\in S\cap V(Y_0)} d(x_i,x_j) + \sum_{x_i\in S\cap V(Y_0)}\sum_{z\in B\cap Z_0}  d(x_i,z)b(z)
+ \sum_{x_i\in S\cap V(Y_0)} d(x_i,t_0) \cdot \left( 2|Y_1| + \sum_{z\in B\cap Z_1} b(z) \right)$$

$$f(S_1)=\sum_{x_i,x_j\in S\cap V(Y_1)} d(x_i,x_j) + \sum_{x_i\in S\cap V(Y_1)}\sum_{z\in B\cap Z_1}  d(x_i,z)b(z)
+ \sum_{x_i\in S\cap V(Y_1)} d(x_i,t_1) \cdot \left( 2|Y_0| + \sum_{z\in B\cap Z_0} b(z) \right)$$

Hence, by adding $t_s$ to $B\cap Z_s=B_s$, with weight equal to $b_s=2|Y_{1-s}| + \sum_{z\in B\cap Z_{1-s}} b(z)$, 
$f(S)$ can be reduced to $f(S_0)$ and $f(S_1)$.
\end{proof}

\section{The algorithm for finding useful edges}
\label{app:three}

\begin{algorithm}[h]
\begin{\algsize}
\KwIn{An instance $G, \{X_1,\ldots, X_k\}, B, b$}
\KwOut{A useful edge, or null}
Pick a node $u$ of the set $X_i$ of the instance  $G, \{X_1,\ldots, X_k\}, B, b$\\
Mark all the nodes as unseen\\
$\dfs[] \gets -1$, \quad $\visited \gets 0$, \quad  $\usefulEdgeFound \gets false$,\quad $\usefulEdge \gets null$\\
\textsc{DFS}($u$,$-1$)\\
\eIf{$\usefulEdgeFound$} {
	\Return{$\usefulEdge$} }{
	\Return{null}
}
\end{\algsize}
\caption{\textsc{UsefulEdge}}
\label{alg:useful}
\end{algorithm}

\begin{algorithm}[h]
\begin{\algsize}
\KwIn{A node $u$, its parent $p$}
\KwOut{A pair ($t$,$Y$), where $t$ is an integer and $Y$ is a set of indices}
\lIf{$\usefulEdgeFound$}{
	\Return{null}
}
Mark $u$ as seen\\ 
$\dfs[u] \gets \visited$\\
$\visited \gets \visited+1$\\
$\furthestAncestor \gets \visited$\\
$Y \gets \emptyset$\\
\lIf{$t \in X_i$}{
	$Y \gets Y \cup \{i\}$
} 
\For{$v\in N(u)$ s.t. $w\neq p$}{
	\eIf{$v$ is unseen} {
		$(t',Y') \gets \mathrm{DFS}( v, u )$\\
		\If{$t'>\dfs[u]$ and $\emptyset\neq Y'\neq \{1,\dots,k\}$} {
			$\usefulEdgeFound \gets true$\\
			$\usefulEdge \gets (u,v)$\\
			\Return{null}
		} 
		$\furthestAncestor \gets \min(\furthestAncestor,t')$\\
		$Y \gets Y \cup Y'$\\
	} {
		$\furthestAncestor \gets \min(\furthestAncestor,\dfs[v])$\\
	}
}
\Return{$(\furthestAncestor,Y)$}
\caption{DFS}
\label{alg:useful-dfs}
\end{\algsize}
\end{algorithm}

\end{document}